\documentclass[twocolumn,english,prl,amssymb,aps,superscriptaddress,showpacs,twocolumn,amsmath,showkeys,floatfix]{revtex4-1}
\usepackage[T1]{fontenc}
\usepackage[latin9]{inputenc}
\usepackage{geometry}
\usepackage[active]{srcltx}
\usepackage{amsmath}
\usepackage{graphicx}
\usepackage{esint}
\usepackage{epsfig}
\usepackage{epstopdf}

\makeatletter
\usepackage{babel}

\makeatother

\usepackage{babel}
\begin{document}

\title{Coherent Population Oscillation-Based Light Storage}

\author{P. Neveu}
\author{ M.-A. Maynard}
\author{R. Bouchez}
\author{J. Lugani}
\address{Laboratoire Aim\'e Cotton, Universit\'e Paris-Sud, ENS Paris-Saclay,
CNRS, Universit\'e Paris-Saclay, 91405 Orsay, France}
\author{R. Ghosh}
\address{Shiv Nadar University, Gautam Budh Nagar, Uttar Pradesh
201314, India}
\author{F. Bretenaker}
\author{F. Goldfarb}
\author{E. Brion}
\address{Laboratoire Aim\'e Cotton, Universit\'e Paris-Sud, ENS Paris-Saclay,
CNRS, Universit\'e Paris-Saclay, 91405 Orsay, France}

\begin{abstract}
We theoretically study the propagation and storage of a classical
field in a $\Lambda$-type atomic medium using coherent population
oscillations (CPOs). We show that the propagation eigenmodes strongly
relate to the different CPO modes of the system. Light storage in
such modes is discussed by introducing a ``populariton'' quantity, a
mixture of populations and field, by analogy to the dark state polariton
used in the context of electromagnetically induced transparency light
storage protocol. As experimentally shown, this memory relies on populations
and is then -- by contrast with usual Raman coherence optical storage
protocols -- robust to dephasing effects.
\end{abstract}

\pacs{42.50.Gy, 42.25.Bs, 42.50.Md}

\maketitle

The architectures proposed to implement optical quantum information
and communication protocols generally rely on quantum memories, \emph{i.e}.
devices able to store quantum states of light and retrieve them on
demand with high fidelity and efficiency \cite{Heshamia2016}. Within
the last decade, much effort has been put towards their implementation
in solid-state systems, ion or neutral atomic ensembles. In this context,
$\Lambda$-type three-level atomic systems have received particular
attention since the coherence between the ground states may have a
long lifetime and can, thus, be used for storage \cite{Gorshkov2006,Lvovsky2009}.
In gas cells, high efficiencies were obtained in alkali-metal atoms
\cite{Novikova2012} using electromagnetically induced transparency
(EIT) close to \cite{Phillips2001} or far off optical resonance \cite{Reim2010},
gradient echo memories \cite{Hetet2008}, or four-wave mixing \cite{Camacho2009}.
Since all these methods are based on the excitation of the Raman coherence
between the lower states of the system, they are sensitive to decoherence
effects. Recently, it was experimentally shown that coherent population
oscillations (CPOs) can be used as a storage medium for light. Experimental
demonstrations were performed using metastable helium (He{*}) vapor
at room temperature \cite{Maynard2014}, as well as in cold and warm
cesium \cite{DeAlmeida2014,Tabosa2015}. CPOs occur in a two-level
system when two detuned coherent electric fields of different amplitudes
drive the same transition. When the detuning between the fields is
smaller than the decay rate of the upper level, the dynamics of the
saturation opens a transparency window in the absorption profile of
the weak field \cite{Boyd1981,Chang2004,Wilson2002}.
The CPO resonance width may be decreased when the upper level decays to
a long-lived shelving state, leading to an ultranarrow CPO resonance
and a memory behavior \cite{Eilam2010}. Another option is to use a $\Lambda$-system
where two CPOs may occur in opposite phase on the two transitions,
leading to a global CPO between the two lower states \cite{Laupretre2012}.
This implies an ultranarrow transmission resonance for the weak field
broadened by the ground states' decay rate, which can be used for
storage \cite{Maynard2014,DeAlmeida2014,Tabosa2015}. Since it involves
only populations, CPO-based light storage protocol is robust to dephasing
effects, by contrast with the EIT-based protocol which involves Raman
coherence. In this Letter, we theoretically explore the $\Lambda$-system
option. First, we study the propagation of a weak signal field in the
medium. We identify eigenmodes of propagation, compute their group
velocities and transmission coefficients, and show that they relate
to different CPO modes. Then, we introduce a new quantity that we
call ``populariton'', by analogy to the dark state polariton (DSP)
put forward in EIT-storage protocols \cite{Fleischhauer2000}, which
allows us to qualitatively understand CPO-based light storage sequence.

\begin{figure}
\epsfig{file=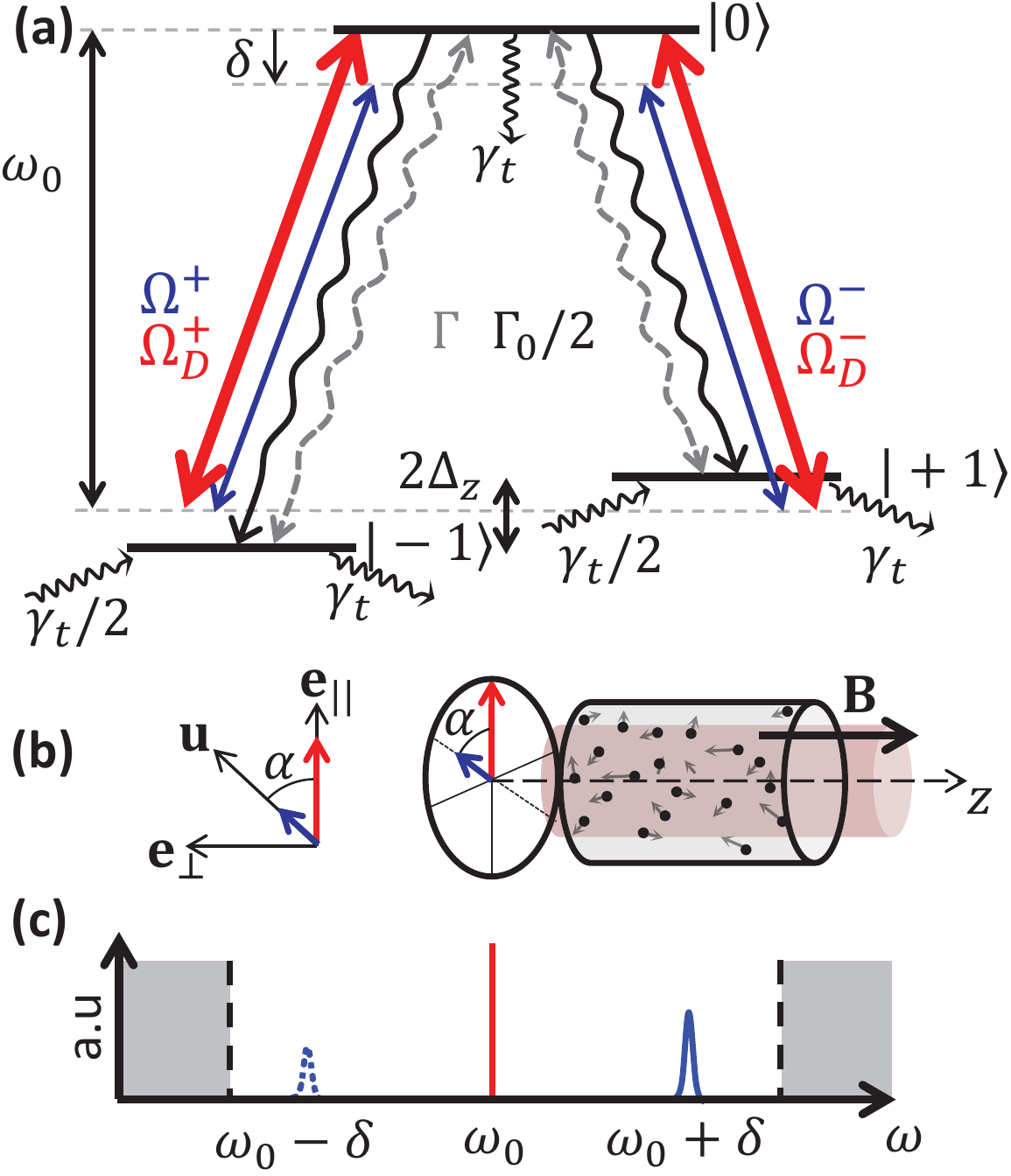,width=0.8\columnwidth}
\caption{(a)\textbf{ }The\textbf{ }$\Lambda$ system of interest. The two circularly
polarized transitions are characterized by the same population (optical coherence) decay
rate $\Gamma_{0}/2$ ($\Gamma$). Atomic thermal motion of results in a transit-population loss of rate $\gamma_{t}\,\left(\ll\Gamma_{0},\Gamma\right)$ for the three states, while the two ground states are fed with the same rate $\gamma_{t}/2$.
The system is coupled to the coherent superposition of two linearly polarized fields,
respectively, along $\left\{ \mathbf{e_{||}},\mathbf{u}\right\}$.
A magnetic field $\mathbf{B}$ Zeeman shifts the two ground states
by $2\Delta_{z}$. (b) The incident
fields propagate along $z$ in the medium. Atoms in the cell interact with the fields within a volume represented in light color, symbolizing
the spatial extension of the beams. (c) Spectrum of the resonant driving
field ($\omega_{0}$) and an example of detuned signal field at ($\omega_{0}+\delta$).
The spectrum of the signal is assumed to be contained within the saturation-broadened
linewidth of the CPO resonance, limited by dashed vertical lines. In
that case, an idler field at ($\omega_{0}-\delta$) is generated along
propagation.\label{System}}
\end{figure}

We consider a $\Lambda$-system similar to the one
which was used to experimentally demonstrate CPO-based light storage,
\emph{i.e.} He{*} at room temperature \cite{Maynard2014}, shown on
Fig. \ref{System}a. Two ground-states Zeeman sublevels $\left|\pm1\right\rangle $
couple to the same excited level $\left|0\right\rangle $ via $\sigma_{\mp}$-polarized
transitions, respectively. $\Gamma_{0}$ denotes the total spontaneous
decay rate from the excited state and $\Gamma\,\left(\gg\Gamma_{0}\right)$
is the common value of the decay rates of the optical coherences $\rho_{0,\pm1}$.
Atomic motion in the vapor cell results in a transit-induced population
loss affecting all states with the same rate $\gamma_{t}\,\left(\ll\Gamma_{0},\Gamma\right)$
and a transit-induced population feeding of rate $\gamma_t/2$ for both ground states (see
Fig. \ref{System}b).

An intense linearly polarized driving field $\mathbf{E_{D}}=\mathcal{E}_{D}e^{-\mathrm{i}\omega_{0}\left(t-z/c\right)}\mathbf{e}_{\mathbf{||}}+\mbox{c.c.}$
and a weak linearly polarized signal field $\mathbf{E}=\mathcal{E}\left(t\right)e^{-\mathrm{i}\omega_{0}\left(t-z/c\right)}\mathbf{u}+\mbox{c.c.}$
are simultaneously sent onto the system. The driving field resonantly
excites the optical transition and $\mathcal{E}_{D}$ is real
positive. The spectrum of the weak time-dependent signal field $\left|\mathcal{E}\left(t\right)\right|\ll\left|\mathcal{E}_{D}\right|$
is assumed to be contained within the driving-field-induced saturation-broadened
linewidth of the CPO resonance. The angle $\alpha$ is defined by $\mathbf{e_{||}}\cdot\mathbf{u}=\cos\alpha$
(see Fig. \ref{System}b), so that the fields
in the circular polarization basis $\mathbf{e_{\pm}}\equiv\frac{\mathbf{e_{||}}\pm\text{i}\mathbf{e_{\perp}}}{\sqrt{2}}$
write 
\begin{align}
\mathbf{E_{D}} & =\frac{\mathcal{E}_{D}}{\sqrt{2}}\left(\begin{array}{c}
1\\
1
\end{array}\right)_{\left\{ \mathbf{\sigma_{+}},\mathbf{\sigma_{-}}\right\} }e^{-\mathrm{i}\omega_{0}\left(t-z/c\right)}+\mbox{c.c.}\label{DefDriving}\\
\mathbf{E} & =\frac{\mathcal{E}\left(t\right)}{\sqrt{2}}\left(\begin{array}{c}
e^{-\mathrm{i}\alpha}\\
e^{\mathrm{i}\alpha}
\end{array}\right)_{\left\{ \mathbf{\sigma_{+}},\mathbf{\sigma_{-}}\right\} }e^{-\mathrm{i}\omega_{0}\left(t-z/c\right)}+\mbox{c.c.}\label{DefProbe}
\end{align}
A static magnetic field is applied along the propagation axis
to Zeeman shift the ground states by
2$\Delta_{z}$, larger than the saturation-broadened linewidth of the
EIT resonances. Thus, Raman
coherent processes between $\left|+1\right\rangle $ and $\left|-1\right\rangle $
can be discarded and the corresponding coherence will be neglected,
\emph{i.e.} $\rho_{1-1}\approx0$.

Let us start with a qualitative discussion of the phenomena at work
in the system. First, we consider the behavior of a single atom subject
to the resonant driving field and a detuned signal field at ($\omega_{0}+\delta$),
typically used in CPO experiments (see Fig. \ref{System}c). The total
intensities $I_{\pm}$ of the $\sigma_{\pm}$ components, which drive
the $\left|\mp1\right\rangle \leftrightarrow\left|e\right\rangle $
transitions, respectively, are modulated at frequency $\delta$. The
atom, therefore, undergoes simultaneous CPOs on the two arms of the
$\Lambda$ system. In particular, when $\alpha=0$, \emph{i.e.} the
two fields have the same polarization, $I_{+}$ and $I_{-}$ oscillate
in phase and the two CPOs combine, leading to a global CPO between
both lower states and the upper one, damped with the rate $\Gamma_{0}$.
Conversely, when $\alpha=\pi/2$, \emph{i.e.} the fields have orthogonal
polarizations, $I_{+}$ and $I_{-}$ oscillate in opposite phase and
the two CPOs are now in antiphase, yielding to an effective CPO between
the two ground states, while the upper state population remains constant
\cite{Laupretre2012}. Thus, this CPO is damped by the ground-state
decay with the rate $\gamma_{t}\,\left(\ll\Gamma_{0}\right)$. The
optical response of the whole medium results from the superposition
of the individual nonlinear behaviors of all the atoms interacting
with the fields; the driving field gets absorbed and a weak so-called
\emph{idler} field at frequency ($\omega_{0}-\delta$), symmetric
of the input signal frequency with respect to $\omega_{0}$, appears
(see Fig. \ref{System}c) \cite{Boyd1981}. Therefore, the output
signal field -- superposition of the distorted input signal and the
generated idler field -- strongly differs from the input one. In the
next paragraphs, we look for the propagation eigenfields, \emph{i.e.}
the signal fields which conserve their polarization and spectrum throughout
propagation. We show that such fields are strongly related to the
CPO modes discussed in this paragraph and, in particular,
have a symmetric spectrum centered at $\omega_{0}$. Moreover, we establish
the analytic expressions of their transmission coefficients and group
velocities.

We describe the dynamics of the system by the set of Maxwell-Bloch
equations perturbatively expanded with respect to the signal field,
in the usual slowly varying envelope approximation for the fields,
and rotating wave approximation (RWA) for the atomic variables expressed
in the frame rotating at $\omega_{0}$. The zeroth order is described
by the following steady-state equations 

\begin{align}
\partial_{z}\Omega_{D}^{\pm}\left(z\right) & =\mathrm{i}\eta\tilde{\rho}_{e\mp1}^{(0)}\left(z\right)\label{PropD}\\
0 & =\left[\hat{H}_{0},\rho^{(0)}\left(z\right)\right]+{\cal D}\left(\rho^{(0)}\left(z\right)\right)\label{OBE0}
\end{align}
where the unperturbed Hamiltonian of the atomic system $\hat{H}_{0}$
includes the internal level structure and interaction with the driving
field, $\Omega_{D}^{\pm}$ denote the Rabi frequencies of the $\sigma_{\pm}$ components
of the driving field, $\tilde{\rho}_{e\pm1}^{(0)}$ denote the zeroth-order
steady-state optical coherences, $\eta$ is the atom field coupling
coefficient and ${\cal D}$ is the operator accounting for spontaneous
emission, dephasing, transit losses and feeding \cite{Supplemental Material}.

At first order, the density matrix $\rho^{(1)}$ obeys 
\begin{align}
\mathrm{i}\hbar\partial_{t}\rho^{(1)} & =\left[\hat{H}_{0},\rho^{(1)}\right]+\left[\hat{H},\rho^{(0)}\right]+{\cal D}\left(\rho^{(1)}\right)\label{OBE1}
\end{align}
where $\Omega^{\pm}$ denote the Rabi frequencies of the $\sigma_{\pm}$ components
of the signal field, $\hat{H}$ is the RWA Hamiltonian describing
the interaction with the signal field. Since we assumed a slowly varying
signal field amplitude ${\cal E}\left(t\right)$ -- the spectrum of
which is included in the saturation-broadened linewidth of the CPO resonance
--, first-order quantities in Eq. (\ref{OBE1}) can be adiabatically
expanded at first order in $\partial_{t}$. We qualitatively explained
above that the weak signal field makes atoms undergo two CPOs on each
arm of the system, which can combine either in phase ($\alpha=0$)
or in opposite phase ($\alpha=\pi/2$). In the former case (symmetric
CPO mode), the first-order ground-state populations are always equal,
thus $\rho_{\Delta}^{(1)}\equiv\rho_{11}^{(1)}-\rho_{-1-1}^{(1)}=0$
while $\rho_{\Sigma}^{(1)}\equiv\rho_{11}^{(1)}+\rho_{-1-1}^{(1)}\neq0$.
In the latter case (antisymmetric CPO mode), we conversely have $\rho_{\Delta}^{(1)}\neq0$
and $\rho_{\Sigma}^{(1)}=0$. In the generic case, Eqs. (\ref{OBE0},\ref{OBE1}) yield

\begin{align}
\rho_{\Delta}^{(1)} & =\frac{-2\beta_{\Delta}}{\left(1+s\right)}\left[1+\left(\frac{1}{2\Gamma}-\frac{\beta_{\Delta}\Gamma}{\left|\Omega_{D}\right|^{2}}\right)\partial_{t}\right]\frac{\Im\mathrm{m}\Big[\Omega^{\perp}\Big]}{\left|\Omega_{D}\right|}\label{eq:DiffPopSommeChamps}\\
\rho_{\Sigma}^{(1)} & =\frac{-2\beta_{\Sigma}}{3\left(1+s\right)}\left[1+\left(\frac{1}{2\Gamma}-\frac{\beta_{\Sigma}\Gamma}{3\left|\Omega_{D}\right|^{2}}\right)\partial_{t}\right]\frac{\Re\mathrm{e}\Big[\Omega^{||}\Big]}{\left|\Omega_{D}\right|}\label{eq:SommePopDiffChamps}
\end{align}
where we introduced the signal field Rabi frequencies components in
the ($\mathbf{e_{||}},\mathbf{e_{\perp}}$) basis $\Omega^{||}\equiv\left[\Omega^{+}+\Omega^{-}\right]/\sqrt{2}$
and $\Omega^{\perp}\equiv\left[\Omega^{+}-\Omega^{-}\right]/\mathrm{i}\sqrt{2}$
(Fig. \ref{System}.b), $\Omega_D$ is the total Rabi frequency of the driving field, $s\equiv3|\Omega_{D}|^{2}/\Gamma_{0}\Gamma$
is the saturation parameter of the transitions and the coefficients
$\beta_{\Delta}\equiv s/\left(3\gamma_{t}/\Gamma_{0}+s\right)$, $\beta_{\Sigma}\equiv s/\left(1+s\right)$
verify $0\leq\beta_{\Delta,\Sigma}\leq1$. The signal field component
$\Omega^{\perp}$ ($\Omega^{||}$), hence, plays the role of a source
term for the population difference $\rho_{\Delta}^{(1)}$ (sum $\rho_{\Sigma}^{(1)}$).
We note that, as the Raman coherence follows the signal field excitation
in an EIT configuration \cite{Fleischhauer2000}, here the sum and
difference of the ground-state populations follow the specific quadratures
$Q^{\perp}\equiv\Im\mathrm{m}\Big[\Omega^{\perp}\Big]$ and $P^{||}\equiv\Re\mathrm{e}\Big[\Omega^{||}\Big]$
of the signal field respectively. The complete description of the
signal field requires the extra two quadratures $Q^{||}\equiv\Im\mathrm{m}\Big[\Omega^{||}\Big]$
and $P^{\perp}\equiv\Re\mathrm{e}\Big[\Omega^{\perp}\Big]$ that we
formally gather with the previous ones in the vector ${\cal S}=\left(\begin{array}{cccc}
P^{\perp}, & P^{||}, & Q^{\perp}, & Q^{||}\end{array}\right)^{\mathrm{T}}$. To determine how ${\cal S}$ propagates, we Fourier transform (FT) the
propagation equation for the first-order field 
\begin{equation}
\left(c\partial_{z}+\text{i}\omega\right)\Omega^{\pm}\left(z,\omega\right)=\mathrm{i}c\eta\tilde{\rho}_{e\mp1}^{(1)}\left(z,\omega\right)\label{eq:PropS}
\end{equation}
as well as Eq. (\ref{OBE1}). Performing a first-order expansion in
$\omega$ -- corresponding to first-order adiabatic expansion in $\partial_{t}$
--, we get \cite{Supplemental Material} 
\begin{equation}
\mathrm{FT}\left[{\cal S}\left(z,t\right)\right]\left(\omega\right)=e^{\int_{0}^{z}{\cal T}\left(\xi\right)\mathrm{d}\xi}\times\mathrm{FT}\left[{\cal S}\left(0,t\right)\right]\left(\omega\right)\label{eq:Prop}
\end{equation}
where ${\cal T}\left(z\right)$ is the diagonal transfer matrix
\begin{equation}
{\cal T}\left(z\right)=-g\mathbb{I}+\left(\begin{array}{cccc}
\mathrm{i}\frac{\omega}{v_{1}} & 0 & 0 & 0\\
0 & 2\beta_{\Sigma}g+\mathrm{i}\frac{\omega}{v_{2}} & 0 & 0\\
0 & 0 & 2\beta_{\Delta}g+\mathrm{i}\frac{\omega}{v_{3}} & 0\\
0 & 0 & 0 & \mathrm{i}\frac{\omega}{v_{1}}
\end{array}\right),\label{eq:T}
\end{equation}
$g=\eta/2\Gamma\left(1+s\right)$ is the absorption coefficient of
the system saturated by the driving field and $v_{i}$'s are group
velocities
\[
\begin{array}{rcl}
v_{1} & = & \frac{c}{1-\frac{c\eta}{2\Gamma^{2}}\cdot\frac{1}{1+s}},\\
v_{2} & = & \frac{c}{1+\frac{c\eta}{2\Gamma^{2}}\cdot\frac{1}{1+s}\cdot\left[2\beta_{\Sigma}^{2}\frac{\Gamma}{s\Gamma_{0}}-\beta_{\Sigma}-1\right]},\\
v_{3} & = & \frac{c}{1+\frac{c\eta}{2\Gamma^{2}}\cdot\frac{1}{1+s}\left[6\beta_{\Delta}^{2}\frac{\Gamma}{s\Gamma_{0}}-\beta_{\Delta}-1\right]}.
\end{array}
\]

\begin{figure}
\epsfig{file=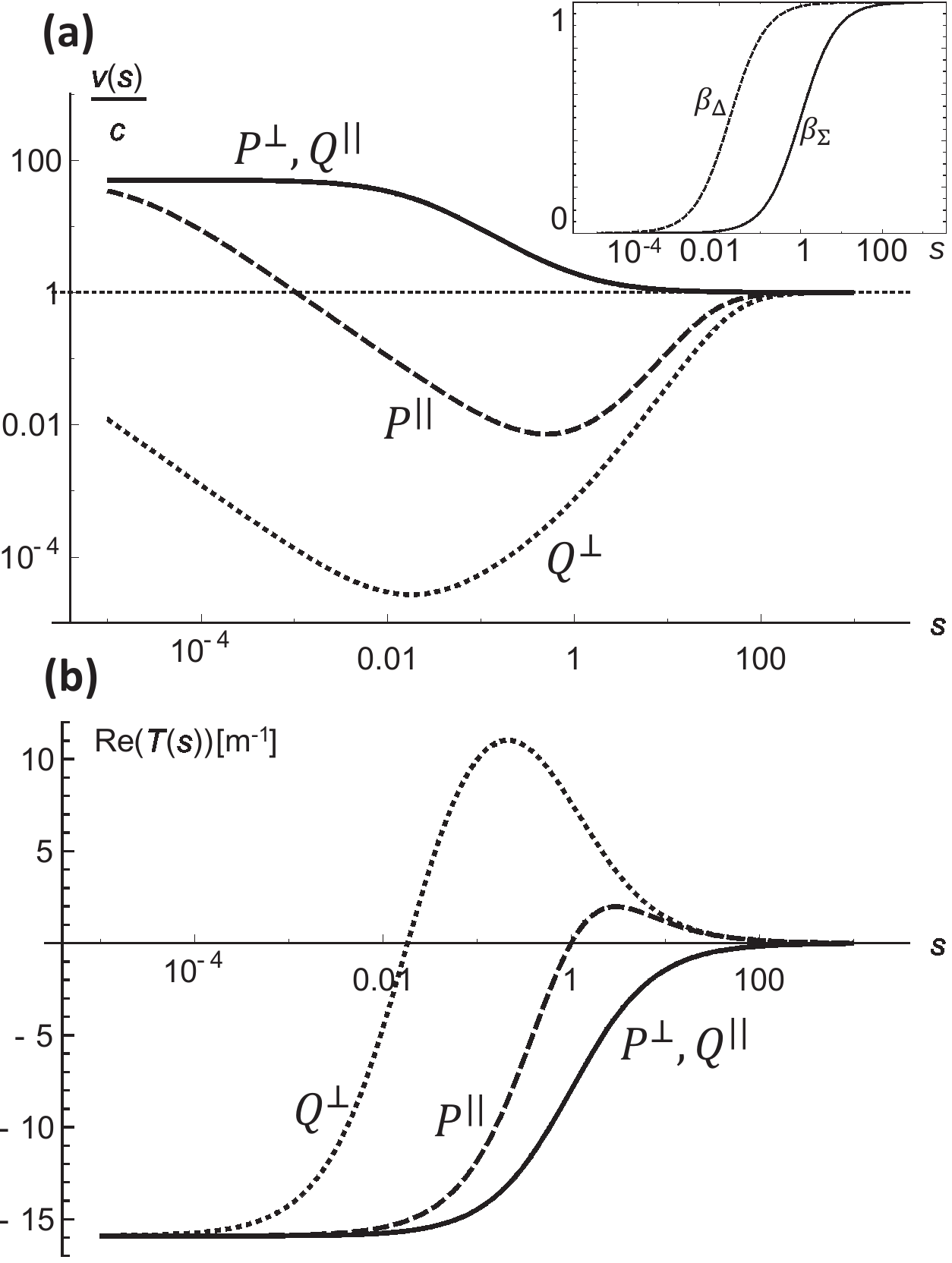,width=0.8\columnwidth}
\caption{Group velocities (a) and transmission coefficients (b) of the eigenquadratures
as functions of the saturation parameter $s$, for He{*} parameters
taken from \cite{Maynard2014}. The inset shows the $\beta_{\Delta,\Sigma}$
parameters as functions of $s$. The quadratures $P^{\perp}$ and
$Q^{||}$ -- which do not explicitly couple to the CPO modes
-- always experience absorption and propagate at a supraluminal group
velocity. By contrast, the quadrature $Q^{\perp}$ ($P^{||}$) which couples to the antisymmetric (symmetric) CPO mode experiences a strongly reduced group velocity and amplification in the regime $3\gamma_t/\Gamma_0\sim0.01<s<100$ ($1<s<100$). Above, the signal cannot interact with the over-saturated atomic system. Below, the states decay ($\beta_{\Sigma,\Delta}\simeq0$) does not allow for CPO.\label{VgGain}}
\end{figure}

Fig. \ref{VgGain} displays the group velocities, transmission coefficients
and $\beta_{\Delta,\Sigma}$ coefficients as functions of the saturation
parameter $s$, obtained with He{*} parameters at room temperature
taken from \cite{Maynard2014}, \emph{i.e. }$\left|\pm1\right\rangle \equiv\left|2^{3}S_{1},m_{J}=\pm1\right\rangle $,
$\left|0\right\rangle \equiv\left|2^{3}P_{1},m_{J}=0\right\rangle $,
$\Gamma/\Gamma_{0}\sim5\cdot10^{2}$, $\gamma_{t}/\Gamma_{0}\sim10^{-2}$,
$\frac{\eta c}{2\Gamma^{2}}\sim1$. Here the optical coherence decay
rate $\Gamma$ is replaced by the Doppler width \cite{Goldfarb2008,Figueroa2006}.
One roughly observes three different regimes. When $s>100$, atoms
are completely saturated by the driving field and the signal field
propagates as in a vacuum. By contrast when $s<0.01$, the linear absorption
regime ($\beta_{\Delta,\Sigma}\thickapprox0$) does not allow for
CPO, the signal field then merely experiences absorption. In between,
the propagation features of the signal field strongly depend on the
driving field intensity. In particular, the quadratures $Q^{\perp}$
and $P^{||}$, which explicitly couple to the CPO modes
via Eqs. (\ref{eq:DiffPopSommeChamps}, \ref{eq:SommePopDiffChamps}),
are amplified and propagate at a strongly reduced group velocity.
By contrast, $P^{\perp}$ and $Q^{||}$, which do not explicitly
couple to CPO modes, always experience absorption and a
supraluminal group velocity.

From Eqs. (\ref{eq:Prop},\ref{eq:T}) we deduce that the input signal
${\cal S}$ is an eigenmode provided that it has a single nonvanishing
quadrature in the basis ($\mathbf{e}_{||},\mathbf{e}_{\perp}$); a propagation eigenmode is linearly polarized along $\mathbf{e}_{||}$
($\alpha=0$) or $\mathbf{e}_{\perp}$ ($\alpha=\pi/2$), and its
Rabi frequency is either real or imaginary, which implies its spectrum
must be symmetric with respect to $\omega_{0}$.

\begin{figure}
\epsfig{file=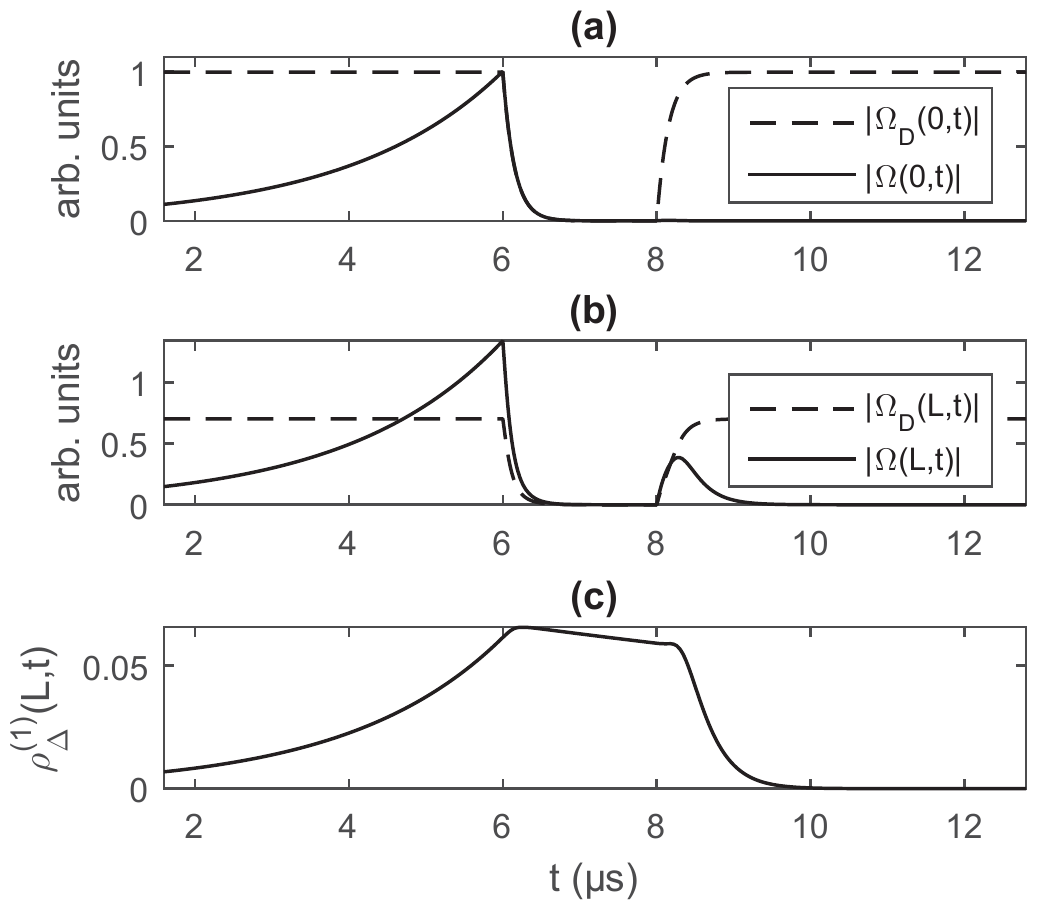,width=1\columnwidth}
\caption{Storage sequence. The input signal field --here at
frequency $\omega_0$-- couples to the population difference
via $Q^\perp$. Renormalized amplitudes of the signal and driving
fields at the entrance (a) and the
exit (b) of the cell, and population difference (c) at the exit of the cell, as functions of time. During the writing step,
the driving field is on, while the signal slowly increases with
a rising exponential shape. The shape of the signal is imprinted
on the population difference. Although the group velocity is strongly
reduced, one observes a leakage of the signal, which was amplified
in the cell. Suddenly, the fields are switched off and the storage
starts. During this period, the generated population difference decays
at rate $\gamma_{t}$. After a 2 $\text{\ensuremath{\mu}}$s storage
time, the driving field is switched on again and a retrieval pulse
of signal exits the cell.\label{fig:StorageSequence}}
\end{figure}

Now, let us consider the specific case of an eigenfield polarized along
$\mathbf{e}_{\perp}$ characterized by ${\cal S}\left(0,t\right)=\left(\begin{array}{cccc}
0 & 0 & \Omega\left(t\right) & 0\end{array}\right)^{\mathrm{T}}$, which propagates with the group velocity $v_{3}$ and couples to
the ground-state population difference (antisymmetric CPO mode). We consider a typical
three-step sequence, used for EIT of CPO storage. The plots displayed
in Fig. \ref{fig:StorageSequence} result from the complete nonperturbative
numerical simulation of Maxwell-Bloch equations with He* parameters taken from \cite{Maynard2014}, in a 6 cm-long cell, with $s\simeq0.1$
so that $\beta_{\Delta}=1$ and $\beta_{\Sigma}=0$. Initially the
driving field is on and the signal field slowly increases. The
saturation parameter is chosen such that $v_{3}\ll c$ in order to
compress the signal field envelope in the medium. At $t=6\,\mathrm{\mu}$s,
the fields are then abruptly switched off. After an arbitrary storage
time (here 2$\,\mathrm{\mu}$s), the driving field is switched on
again and a retrieved pulse exits the cell.

In the same way as the quadrature $Q^{\perp}$ is a
source term for the population difference $\rho_{\Delta}^{(1)}$ in
Eq. (\ref{eq:DiffPopSommeChamps}), $\rho_{\Delta}^{(1)}$ conversely
appears as a driving term in the following propagation equation of
$Q^{\perp}$ 
\begin{equation}
\left(c\partial_{z}+\partial_{t}-cg\right)Q^{\perp}=-\frac{\eta c}{2|\Omega_{D}|}\partial_{t}\rho_{\Delta}^{(1)}\label{eq:PropagCoupl}
\end{equation}
These relations are reminiscent of those one can write for the Raman
coherence and the field in an EIT configuration. Thus, by analogy
with the DSP picture \cite{Fleischhauer2000}, we define\textbf{ }a
new quantity, superposition of the quadrature $Q^{\perp}$\textbf{
}and the population difference $\rho_{\Delta}^{(1)}$, the \emph{populariton}

\begin{equation}
{\cal P}=\cos\left(\Theta\right)Q^{\perp}-\sqrt{\frac{\eta c}{8}}\sin\left(\Theta\right)\rho_{\Delta}^{(1)}\label{eq:Populariton}
\end{equation}
with the mixing angle $\Theta$ defined by $\tan\Theta=\sqrt{\frac{\eta c}{2|\Omega_{D}|^{2}}}$,
controlled by the driving field intensity. This quantity has light and matter components during the writing and retrieval steps
($0<\Theta<\frac{\pi}{2}$), but is exclusively in the form of the
difference of populations during the storage step ($\Theta=\frac{\pi}{2}$).
Using Eqs. (\ref{eq:DiffPopSommeChamps}, \ref{eq:PropagCoupl}, \ref{eq:Populariton}),
one can show that $\cos\left(\Theta\right){\cal P}=\left[1-\sin^{2}\left(\Theta\right)\frac{\Gamma}{\left|\Omega_{D}\right|^{2}}\partial_{t}\right]Q^{\perp}$
and $\sin\left(\Theta\right){\cal P}=-\sqrt{\frac{\eta c}{8}}\left[1+\cos^{2}\left(\Theta\right)\frac{\Gamma}{\left|\Omega_{D}\right|^{2}}\partial_{t}\right]\rho_{\Delta}^{(1)}$,
which, together with Eq. (\ref{eq:PropagCoupl}) lead to the propagation
equation for the populariton
\begin{equation}
\left(\partial_{z}+\frac{2-\cos^{4}\left(\Theta\right)}{v_{3}\left(\Theta\right)}\partial_{t}\right){\cal P}=g\left(1+\sin^{2}\left(\Theta\right)\right){\cal P}\label{eq:PropaPopula}
\end{equation}
with the group velocity $v_{3}\left(\Theta\right)/2-\cos^{4}\left(\Theta\right)$
and an amplification factor $g\left(1+\sin^{2}\left(\Theta\right)\right)$.
The retrieval process can be interpreted in the same way as in EIT
protocols: when the driving field is switched on again after storage,
${\cal P}$ takes back a signal field component, \emph{i.e.} the retrieved
signal pulse. Moreover the lifetime of the memory corresponds to the
lifetime of ${\cal P}$ during the storage step, \emph{i.e.} the ground-state-population
difference, which decays at rate $\gamma_{t}$.

Since we considered the input signal spectrum to be included within the CPO linewidth,
the first-order adiabatic restriction erases dispersive
effects along propagation and rigorous optimization such as in EIT protocols \cite{GorshkovPRL,GorshkovPRA2,GorshkovPRA4,Phillips2008,Novikova2012} would require to go beyond this approximation. However, as well as EIT occurs when the saturation broadening overcomes the Raman decoherence \cite{Fleischhauer2005}, our adiabatic model shows that CPO occurs when the saturation broadening overcomes the ground-states decay ($s>3\gamma_t/\Gamma_0$). As for EIT-based storage \cite{Fleischhauer2005,Phillips2008}, optimal efficiencies are expected for abrupt switching of the driving field and moderate optical depth \cite{Supplemental Material}.

Above, we considered that the signal field is an eigenvector of the
transfer matrix ${\cal T}$. For an arbitrary linearly polarized signal
field with an arbitrary spectrum, the populariton picture can still
be used for the storage of the $Q^{\perp}$ quadrature of the distorted
signal field. The same kind of calculations and interpretation can
actually be done for the other CPO (\emph{i.e.} symmetric) excitation
mode, characterized by the ground-states population sum $\rho_{\Sigma}^{(1)}$,
coupled to the quadrature $P^{||}$, with a lifetime $\Gamma_{0}^{-1}$. 
In that case the broader CPO
linewidth allows for shorter input signal pulses.

In this Letter, we studied the propagation of a weak signal field
in a $\Lambda$-type atomic medium resonantly driven by a strong pump
field. We identified four propagation eigenmodes, two of which directly
couple to the CPO modes of the medium. To interpret CPO-based
light storage in such modes we introduced the populariton, mixture
of light and matter, which is an analogue of the DSP introduced in
\cite{Fleischhauer2000} to interpret EIT-based memory. The main advantage
of the CPO-based memory described here, as experimentally shown \cite{Maynard2014},
is its robustness to dephasing effects since it relies on populations. Our study applies beyond He* warm vapor to any system, e.g. solid-state \cite{Chang2004} or cold atoms  \cite{DeAlmeida2014} where CPO was observed.
Future work will determine whether it can be used to simultaneously store both non-commuting quadratures of a light field.

\begin{acknowledgments}
The authors acknowledge funding by Indo-French CEFIPRA, Labex PALM,
R\'egion IdF DIM Nano-K, IUF. The work of M. A. M. is supported
by the D\'el\'egation G\'en\'erale de l'Armement (DGA).
\end{acknowledgments}

\end{document}


\title{Coherent Population Oscillation-Based Light Storage -- Supplemental
Material}

\author{P. Neveu}
\author{ M.-A. Maynard}
\author{R. Bouchez}
\author{J. Lugani}
\address{Laboratoire Aim\'e Cotton, Universit\'e Paris-Sud, ENS Paris-Saclay,
CNRS, Universit\'e Paris-Saclay, 91405 Orsay, France}
\author{R. Ghosh}
\address{Shiv Nadar University, Gautam Budh Nagar, Uttar Pradesh
201314, India}
\author{F. Bretenaker}
\author{F. Goldfarb}
\author{E. Brion}
\address{Laboratoire Aim\'e Cotton, Universit\'e Paris-Sud, ENS Paris-Saclay,
CNRS, Universit\'e Paris-Saclay, 91405 Orsay, France}

\maketitle
In this Supplemental Material, we give technical details and intermediate steps of the calculations discussed in the main text. In Section A, we provide the perturbative expansion of the optical Bloch equations for our system of interest, and detail the adiabatic expansion for the first order. In Section B, we briefly explain the way we derive the propagation equation for the eigencomponents. In Section C, we give precisions concerning the propagation equation for the populariton. Finally, in Section D, we qualitativaly discuss the influence of the optical depth and the driving field switching sharpness according to our adiabatic model and justify it with full numerical simulations.

\subsection{Pertubative development of the optical Bloch equations (OBE)}

\subsubsection{Zeroth order}

We consider the $\Lambda$-system represented in Fig. 1.a from the
text. Zeroth-order OBE (Eqs. {[}4{]} of the
text) write \foreignlanguage{french}{
\begin{eqnarray*}
\partial_{t}\tilde{\rho}_{e1}^{(0)} & = & -\left[\Gamma-\mathrm{i}\left(\Delta_{D}+\Delta_{z}\right)\right]\tilde{\rho}_{e1}^{(0)}-\mathrm{i}\left(1-2\rho_{11}^{(0)}-\rho_{-1-1}^{(0)}\right)\Omega_{D}^{-}+\mathrm{i}\tilde{\rho}_{-11}^{(0)}\Omega_{D}^{+}\\
\partial_{t}\tilde{\rho}_{e-1}^{(0)} & = & -\left[\Gamma-\mbox{i}\left(\Delta_{D}-\Delta_{z}\right)\right]\tilde{\rho}_{e-1}^{(0)}+\mathrm{i}\tilde{\rho}_{1-1}^{(0)}\Omega_{D}^{-}-\mathrm{i}\left(1-\rho_{11}^{(0)}-2\rho_{-1-1}^{(0)}\right)\Omega_{D}^{+}\\
\partial_{t}\tilde{\rho}_{1-1}^{(0)} & = & -\left[\gamma_{R}+2\mathrm{i}\Delta_{z}\right]\tilde{\rho}_{1-1}^{(0)}+\mbox{i}\left(\tilde{\rho}_{e-1}^{(0)}\Omega_{D}^{-*}-\tilde{\rho}_{e1}^{(0)*}\Omega_{D}^{+}\right)\\
\partial_{t}\rho_{11}^{(0)} & = & -\left(\gamma_{t}+\frac{\Gamma_{0}}{2}\right)\rho_{11}^{(0)}-\frac{\Gamma_{0}}{2}\rho_{-1-1}^{(0)}-\mathrm{i}\left(\tilde{\rho}_{e1}^{(0)*}\Omega_{D}^{-}-\tilde{\rho}_{e1}^{(0)}\Omega_{D}^{-*}\right)+\frac{\Gamma_{0}+\gamma_{t}}{2}\\
\partial_{t}\rho_{-1-1}^{(0)} & = & -\left(\gamma_{t}+\frac{\Gamma_{0}}{2}\right)\rho_{-1-1}^{(0)}-\frac{\Gamma_{0}}{2}\rho_{11}^{(0)}-\mathrm{i}\left(\tilde{\rho}_{e-1}^{(0)*}\Omega_{D}^{+}-\tilde{\rho}_{e-1}^{(0)}\Omega_{D}^{+*}\right)+\frac{\Gamma_{0}+\gamma_{t}}{2}
\end{eqnarray*}
}where $\gamma_{R}$ is the Raman coherence decay rate, $\Delta_{D}$
is the detuning of the driving field from the atomic resonance $\omega_{0}$,
and the other quantities are defined in the text. The convention we
used for the Rabi frequencies is $\hbar\Omega_{D}^{\pm}\equiv d\mathcal{E}_{D}^{\pm}$
with $d$ the common atomic dipole of the transitions. We assume that
the zeroth order can be treated in the steady-state regime ($\partial_{t}^{(0)}=0$)
and that the driving field it resonant (Fig. 1.a of the main text) \emph{i.e.}
$\Delta_{D}=0$. One note that the choice of polarization induces
$\Omega_{D}^{+}=\Omega_{D}^{-}=\left|\Omega_{D}^{\pm}\right|=\left|\Omega_{D}\right|/\sqrt{2}$
. Finally, we consider that the Zeeman shift is large enough to avoid
the Raman coherence to build up while the optical transitions remain driven by the field \emph{i.e.} $\Delta_Z\ll\Gamma$. Then the zeroth order set of OBE
boils down to \foreignlanguage{french}{
\begin{eqnarray*}
\tilde{\rho}_{e1}^{(0)} & =\frac{\mathrm{i}}{2\Gamma\left(1+s\right)}\Omega_{D}^{-}=\tilde{\rho}_{e-1}^{(0)},\quad\mathrm{and}\quad\rho_{\pm1\pm1}^{(0)}=\frac{1/2+s/3}{1+s}
\end{eqnarray*}
}with $s=\frac{3\left|\Omega_{D}\right|^{2}}{\left(\gamma_{t}+\Gamma_{0}\right)\Gamma}\simeq\frac{3\left|\Omega_{D}\right|^{2}}{\Gamma_{0}\Gamma}$
the saturation parameter of the transitions.

\subsubsection{First order -- Temporal point of view}

The complete set (excluding the Raman coherence) of the first-order
OBE (Eqs. (5) of the text), using the previous results and approximations, writes as
follows \foreignlanguage{french}{
\begin{eqnarray*}
\partial_{t}\tilde{\rho}_{e-1}^{(1)} & = & -\Gamma\tilde{\rho}_{e-1}^{(1)}+\frac{\mathrm{i}}{2\left(1+s\right)}\Omega^{+}+\mathrm{i}\left(\rho_{11}^{(1)}+2\rho_{-1-1}^{(1)}\right)\left|\Omega_{D}\right|/\sqrt{2}\\
\partial_{t}\tilde{\rho}_{e1}^{(1)} & = & -\Gamma\tilde{\rho}_{e1}^{(1)}+\frac{\mathrm{i}}{2\left(1+s\right)}\Omega^{-}+\mathrm{i}\left(\rho_{-1-1}^{(1)}+2\rho_{11}^{(1)}\right)\left|\Omega_{D}\right|/\sqrt{2}\\
\partial_{t}\rho_{-1-1}^{(1)} & = & -\left(\gamma_{t}+\frac{\Gamma_{0}}{2}\right)\rho_{-1-1}^{(1)}-\frac{\Gamma_{0}}{2}\rho_{11}^{(1)}-\sqrt{2}\left|\Omega_{D}\right|\Im\mathrm{m}\left(\tilde{\rho}_{e-1}^{(1)}\right)-\frac{\left|\Omega_{D}\right|}{\sqrt{2}\Gamma\left(1+s\right)}\Re\mathrm{e}\left(\Omega^{+}\right)\\
\partial_{t}\rho_{11}^{(1)} & = & -\left(\gamma_{t}+\frac{\Gamma_{0}}{2}\right)\rho_{11}^{(1)}-\rho_{-1-1}^{(1)}\frac{\Gamma_{0}}{2}-\sqrt{2}\left|\Omega_{D}\right|\Im\mathrm{m}\left(\tilde{\rho}_{e1}^{(1)}\right)-\frac{\left|\Omega_{D}\right|}{\sqrt{2}\Gamma\left(1+s\right)}\Re\mathrm{e}\left(\Omega^{-}\right)
\end{eqnarray*}
}We have seen in the text that relevant quantities are the sum and
difference of first-order populations $\rho_{\Delta/\Sigma}^{(1)}$,
which can be written as functions of orthogonal/parallel signal field
components $\Omega^{\perp,||}$ respectively. From the previous system,
one deduces the system (notations are introduced in the text): \foreignlanguage{french}{
\begin{eqnarray*}
\partial_{t}\left(\tilde{\rho}_{e1}^{(1)}+\tilde{\rho}_{e-1}^{(1)}\right) & = & -\Gamma\left(\tilde{\rho}_{e1}^{(1)}+\tilde{\rho}_{e-1}^{(1)}\right)+\frac{\mathrm{i}}{\sqrt{2}\left(1+s\right)}\Omega^{||}+3\mathrm{i}\rho_{\Sigma}^{(1)}\left|\Omega_{D}\right|/\sqrt{2}\\
\partial_{t}\left(\tilde{\rho}_{e1}^{(1)}-\tilde{\rho}_{e-1}^{(1)}\right) & = & -\Gamma\left(\tilde{\rho}_{e1}^{(1)}-\tilde{\rho}_{e-1}^{(1)}\right)+\frac{1}{\sqrt{2}\left(1+s\right)}\Omega^{\perp}+\mathrm{i}\rho_{\Delta}^{(1)}\left|\Omega_{D}\right|/\sqrt{2}\\
\partial_{t}\rho_{\Delta}^{(1)} & = & -\gamma_{t}\rho_{\Delta}^{(1)}-\sqrt{2}\left|\Omega_{D}\right|\Im\mathrm{m}\left(\tilde{\rho}_{e1}^{(1)}-\tilde{\rho}_{e-1}^{(1)}\right)-\frac{\left|\Omega_{D}\right|}{\Gamma\left(1+s\right)}\Im\mathrm{m}\left(\Omega^{\perp}\right)\\
\partial_{t}\rho_{\Sigma}^{(1)} & = & -\left(\gamma_{t}+\Gamma_{0}\right)\rho_{\Sigma}^{(1)}-\sqrt{2}\left|\Omega_{D}\right|\Im\mathrm{m}\left(\tilde{\rho}_{e1}^{(1)}+\tilde{\rho}_{e-1}^{(1)}\right)-\frac{\left|\Omega_{D}\right|}{\Gamma\left(1+s\right)}\Re\mathrm{e}\left(\Omega^{||}\right)
\end{eqnarray*}
}

\subsubsection{First order -- Fourier point of view}

We make the Fourier transform of the previous system with the convention
$\partial_{t}\overset{\mathrm{F.T}}{\longrightarrow}+\mathrm{i}\omega$.
Keeping in mind that $\mathrm{FT}\left[f\left(z,t\right)\right]=f\left(z,\omega\right)\leftrightarrow\mathrm{FT}\left[f^{*}\left(z,t\right)\right]=f^{*}\left(z,-\omega\right)$
and that populations in the temporal domain are real quantities so that $\rho_{ii}\left(z,\omega\right)=\rho_{ii}\left(z,-\omega\right)$,
we get \foreignlanguage{french}{
\begin{eqnarray*}
\left(\tilde{\rho}_{e1}^{(1)}+\tilde{\rho}_{e-1}^{(1)}\right)\left(z,\omega\right) & = & \frac{1}{\Gamma+\mathrm{i}\omega}\left[\frac{\mathrm{i}}{\sqrt{2}\left(1+s\right)}\Omega^{||}+3\mathrm{i}\rho_{\Sigma}^{(1)}\left|\Omega_{D}\right|/\sqrt{2}\right]\left(z,\omega\right)\\
\left(\tilde{\rho}_{e1}^{(1)}-\tilde{\rho}_{e-1}^{(1)}\right)\left(z,\omega\right) & = & \frac{1}{\Gamma+\mathrm{i}\omega}\left[\frac{1}{\sqrt{2}\left(1+s\right)}\Omega^{\perp}+\mathrm{i}\rho_{\Delta}^{(1)}\left|\Omega_{D}\right|/\sqrt{2}\right]\left(z,\omega\right)\\
\rho_{\Delta}^{(1)}\left(z,\omega\right) & = & -\frac{2+\mathrm{i}\frac{\omega}{\Gamma}}{\left(1+\mathrm{i}\frac{\omega}{\gamma_{t}}\right)\left(1+\mathrm{i}\frac{\omega}{\Gamma}\right)+\left|\Omega_{D}\right|^{2}/\gamma_{t}\Gamma}\frac{\left|\Omega_{D}\right|}{1+s}\frac{1}{\gamma_{t}\Gamma}Q^{\perp}\left(z,\omega\right)\\
\rho_{\Sigma}^{(1)}\left(z,\omega\right) & = & -\frac{2+\mathrm{i}\frac{\omega}{\Gamma}}{\left(1+\mathrm{i}\frac{\omega}{\Gamma}\right)\left(1+\mathrm{i}\frac{\omega}{\gamma_{t}+\Gamma_{0}}\right)+3\left|\Omega_{D}\right|^{2}/\Gamma\left(\gamma_{t}+\Gamma_{0}\right)}\frac{\left|\Omega_{D}\right|}{1+s}\frac{1}{\left(\gamma_{t}+\Gamma_{0}\right)\Gamma}P^{||}\left(z,\omega\right)
\end{eqnarray*}
}with
\begin{equation}
Q^{\perp,||}\left(z,\omega\right)=\frac{\Omega^{\perp,||}\left(z,\omega\right)-\Omega^{\perp,||*}\left(z,-\omega\right)}{2\mathrm{i}},\quad\quad P^{\perp,||}\left(z,\omega\right)=\frac{\Omega^{\perp,||}\left(z,\omega\right)+\Omega^{\perp,||*}\left(z,-\omega\right)}{2}\label{eq:Q}
\end{equation}
When adiabatically developed at the first order in $\omega$, the
populations sum and difference write \foreignlanguage{french}{
\begin{eqnarray*}
\rho_{\Delta}^{(1)}\left(z,\omega\right) & = & \frac{-2\beta_{\Delta}}{\left(1+s\right)\left|\Omega_{D}\right|}\left[1+\mathrm{i}\omega\left(\frac{1}{2\Gamma}-\beta_{\Delta}\frac{\Gamma+\gamma_{t}}{\left|\Omega_{D}\right|^{2}}\right)\right]Q^{\perp}\\
\rho_{\Sigma}^{(1)}\left(z,\omega\right) & = & \frac{-2\beta_{\Sigma}}{3\left(1+s\right)\left|\Omega_{D}\right|}\left[1+\mathrm{i}\omega\left(\frac{1}{2\Gamma}-\beta_{\Sigma}\frac{\Gamma+\Gamma_{0}+\gamma_t}{3\left|\Omega_{D}\right|^{2}}\right)\right]P^{||}
\end{eqnarray*}
}which coincide with Eqs. (6,7) in the text with the assumption $\gamma_{t}\ll\Gamma_{0}\ll\Gamma$.

\subsection{Propagation equations}

The propagation equation writes, in the Fourier domain with the convention
$\partial_{t}\overset{\mathrm{F.T}}{\longrightarrow}+\mathrm{i}\omega$:
\begin{equation}
\left(c\partial_{z}+\text{i}\omega\right)\Omega^{\pm}\left(z,\omega\right)=\mathrm{i}c\eta\tilde{\rho}_{e\mp1}^{(1)}\left(z,\omega\right)\label{eq:Prop}
\end{equation}
with $\eta\equiv\frac{n\omega_{0}\left|d\right|^{2}}{2\hbar c\epsilon_{0}}$
the atom-field dipolar coupling coefficient. From the Eqs. (\ref{eq:Q},\ref{eq:Prop}),
we can then obtain the propagation equations for each quadrature:
\[
\begin{array}{rcl}
\partial_{z}Q^{\perp}\left(z,\omega\right) & = & \left[\frac{\eta}{2\Gamma\left(1+s\right)}\left(2\beta_{\Delta}-1\right)-\mathrm{i}\frac{\omega}{c}\left\{ 1+\frac{c\eta}{2\Gamma^{2}\left(1+s\right)}\left[6\beta_{\Delta}^{2}\frac{\Gamma+\gamma_{t}}{s\Gamma_{0}}-\beta_{\Delta}-1\right]\right\} \right]Q^{\perp}\\
\partial_{z}P^{\perp}\left(z,\omega\right) & = & \left[-\frac{\eta}{2\Gamma\left(1+s\right)}-\mathrm{i}\frac{\omega}{c}\left\{ 1-\frac{c\eta}{2\Gamma^{2}\left(1+s\right)}\right\} \right]P^{\perp}\\
\partial_{z}Q^{||}\left(z,\omega\right) & = & \left[-\frac{\eta}{2\Gamma\left(1+s\right)}-\mathrm{i}\frac{\omega}{c}\left\{ 1-\frac{c\eta}{2\Gamma^{2}\left(1+s\right)}\right\} \right]Q^{||}\\
\partial_{z}P^{||}\left(z,\omega\right) & = & \left[\frac{\eta}{2\Gamma\left(1+s\right)}\left(2\beta_{\Sigma}-1\right)-\mathrm{i}\frac{\omega}{c}\left\{ 1+\frac{c\eta}{2\Gamma^{2}\left(1+s\right)}\left[2\beta_{\Sigma}^{2}\frac{\Gamma+\gamma_{t}+\Gamma_{0}}{s\Gamma_{0}}-\beta_{\Sigma}-1\right]\right\} \right]P^{||}
\end{array}
\]
These propagations equations can easily be integrated to obtain Eqs.
(9,10) of the text, assuming $\gamma_{t}\ll\Gamma_{0}\ll\Gamma$.

\subsection{Populariton picture}

\subsubsection{Antisymmetric CPO mode}

In the text, we first considered the storage of the quadrature $Q^{\perp}$,
\emph{i.e.} in the antisymmetric CPO mode. To optimize the efficiency
of the storage, which is minimizing the group velocity and absorption,
we chose the specific value $s\simeq0.1$ which yields a simple propagation
equation for the populariton while keeping all relevant physical features
of the phenomenon. Here, we extend the results presented in the text
by allowing the saturation parameter to take any value within $0.1<s<10$.
In this interval, we define the populariton as follows
\[
{\cal P}=\frac{1}{1+s}\cos\left(\Theta\right)Q^{\perp}-\sqrt{\frac{\eta c}{8}}\sin\left(\Theta\right)\rho_{\Delta}^{(1)}
\]
and we have, in good approximation
\[
\begin{array}{rcl}
v_{3} & = & \frac{c}{1+\frac{c\eta}{2\Gamma^{2}}\cdot\frac{1}{1+s}\left[6\beta_{\Delta}^{2}\frac{\Gamma}{s\Gamma_{0}}-\beta_{\Delta}-1\right]}\\
 & \simeq & \frac{s\left(1+s\right)\Gamma_{0}\Gamma}{3\eta}
\end{array}\qquad\mbox{and}\qquad\begin{array}{rcl}
\rho_{\Delta}^{(1)} & = & \frac{-2}{\left(1+s\right)}\left[1+\mathrm{i}\omega\left(\frac{1}{2\Gamma}-\frac{3}{s\Gamma_{0}}\right)\right]\sqrt{\frac{s\Gamma_{0}\Gamma}{3}}Q^{\perp}\\
 & \simeq & \frac{-2}{\left(1+s\right)}\left[1-\mathrm{i}\omega\frac{3}{s\Gamma_{0}}\right]\sqrt{\frac{s\Gamma_{0}\Gamma}{3}}Q^{\perp}
\end{array}
\]
where we used the fact that, in our system of interest (He{*}), one
has $\Gamma/\Gamma_{0}\sim5\cdot10^{2}$, $\gamma_{t}/\Gamma_{0}\sim10^{-2}$,
$\frac{\eta c}{2\Gamma^{2}}\sim1$. One must take into account the
driving field absorption in the medium through its propagation equation
\[
\partial_{z}\Omega_{D}^{\pm}\left(z\right)=\mathrm{i}\eta\tilde{\rho}_{e\mp1}^{(0)}\left(z\right)\quad\longrightarrow\quad\partial_{z}s=-\frac{\eta}{\Gamma}\frac{s}{1+s}=-\frac{\eta}{\Gamma}\beta_{\Sigma}
\]
Then, from Eqs. (10,11) in the text, taking into account the $z$
dependance of $\Theta$, and remaining at a first order in $\partial_{t}$,
we obtain the propagation equation for the populariton
\[
\partial_{z}{\cal P}=\left[\frac{\eta}{2\Gamma\left(1+s\right)}\left(1+\sin^{2}\Theta+2\beta_{\Sigma}\right)-\mathrm{i}\frac{\omega}{v_{3}}\left(2-\cos^{4}\Theta\right)\right]{\cal P}
\]
which coincides with Eq. (13) in the text with the explicit dependence
in $s$. Note that it is irrelevant to consider the case when $\beta_{\Delta}<1$
because in that case $\rho_{\Delta}^{(1)}$ and $Q^{\perp}$ are not
coupled any more.

\subsubsection{Symmetric CPO mode}

One can define another populariton, for the symmetric CPO mode :

\[
{\cal P}'=\frac{1}{1+s}\cos\left(\Theta\right)P^{||}-3\sqrt{\frac{\eta c}{8}}\sin\left(\Theta\right)\rho_{\Sigma}^{(1)}
\]
The relevant regime for this CPO mode is when $s\sim10$. In that
case
\[
v_{2}=\frac{c}{1+\frac{c\eta}{2\Gamma^{2}}\cdot\frac{1}{1+s}\cdot\left[2\beta_{\Sigma}^{2}\frac{\Gamma}{s\Gamma_{0}}-\beta_{\Sigma}-1\right]}\simeq\frac{s^{2}\Gamma_{0}\Gamma}{\eta}\quad\mbox{and}\quad\rho_{\Sigma}^{(1)}\simeq\frac{-2}{3s}\left[1-\frac{1}{s\Gamma_{0}}\mathrm{i}\omega\right]\sqrt{\frac{s\Gamma_{0}\Gamma}{3}}P^{||}
\]
One can show that the propagation for this populariton is similar
to the previous one
\[
\partial_{z}{\cal P}'=\left[\frac{\eta}{2\Gamma s}\left(3+\sin^{2}\Theta\right)-\mathrm{i}\frac{\omega}{v_{2}}\left(2-\cos^{4}\Theta\right)\right]{\cal P}'
\]

\subsection{Sharpness of the storage step -- Optical depth influence}

A complete description of the optimization of the CPO-based storage protocol is out of the scope of this manuscript. Indeed, the first-order adiabatic development erases dispersive effects along propagation, forbidding any discussion concerning the influence of the temporal shapes of the fields. In EIT-based storage protocols, such optimization has been investigated both theoretically (see Refs. [20-22] of the main text) and experimentally (see Ref. [4,23] of the main text). Besides, what really matters would be the quantum fidelity of CPO-storage, which is not adressed in the manuscript. Nevertheless, one can exploit the adiabatic model to qualitatively discuss the impact of the sharpness of the storage sequence as well as the influence of the optical depth on the classical storage efficiency defined by
\[
e=\frac{\int_{T}^{\infty}\left|{\cal E}\left(L,t\right)\right|^2\mathrm{d}t}{\int_{-\infty}^{\infty}\left|{\cal E}\left(0,t\right)\right|^2\mathrm{d}t}
\]
The CPO dynamics is governed by the saturation of the system induced by the driving field. This implies some basic criteria to actually store a signal using CPO. In the next paragraph, we give qualitative discussions in light of the adiabatic model and justify them with complete numerical simulation of Maxwell-Bloch equations.

\begin{figure}[h]
\epsfig{file=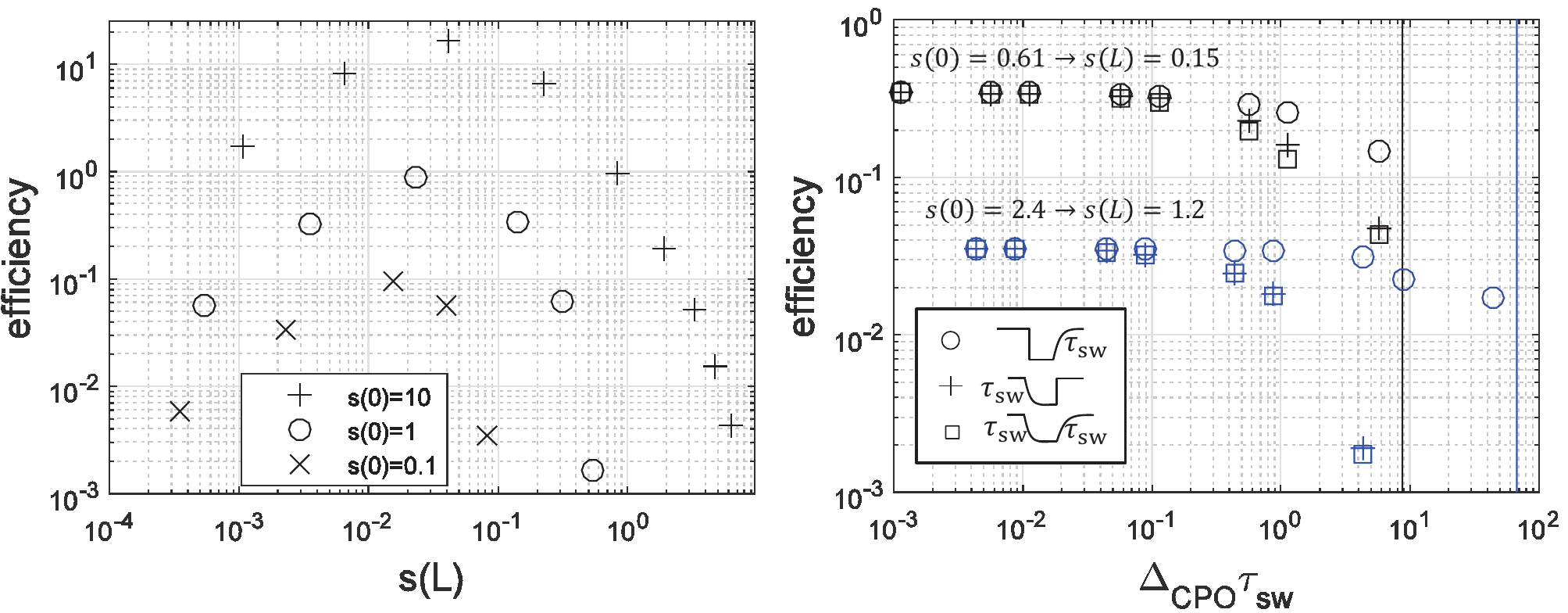,width=1\columnwidth}
\caption{Numerical estimation of the classical storage efficiency of a 2$\,\mu$s exponentially rising signal pulse, for a storage duration of 3$\,\mu$s. (a) Storage efficiency for various optical depths, for different input saturations $s(0)$. The efficiency is optimum when the optical depth is such that the output saturation is $s(L)\sim 3\gamma_t/\Gamma_0$. Efficiency can be larger than one because of the amplification process occuring in the medium. (b) Storage efficiency as a function of the driving field switching time $\tau_{\mathrm{sw}}$ in unit of $1/\Delta_{\mathrm{CPO}}$, for two saturation conditions, in a medium of optical depth 0.6 so that the saturation $s$ slightly decreases along the cell. Vertical lines correspond to the memory lifetime $1/\gamma_t$, which is an upper bound for a relevant driving field sharpness. As expected qualitatively, the sharpness of the storage step has a stronger influence on the efficiency than the retrieval's.\label{Switch}}
\end{figure}

Our framework describes a slowly varying regime (as compared to the timescale given by the saturation-broadened CPO linewidth $\Delta_{\mathrm{CPO}}=\gamma_t+\left|\Omega_D\right|^2/\Gamma$) in which populations follow the signal field. For a weakly-saturated system ($s\ll0.01\sim3\gamma_t/\Gamma_0$), the signal field gets absorbed while propagating at large group velocities (see Fig. 2 of the manuscript). In other terms, in such a regime, the signal field is lost. This low-saturation regime can be reached for two reasons; either the optical depth is too high so that the driving field is strongly absorbed, or, the driving field is switched off too slowly, which ensures that our adiabatic treatment is applicable. Considering the former case (see Fig. \ref{Switch}a), we then expect to have better storage efficiencies when the optical depth is such that, for a given input saturation $s(0)$, the ouput saturation is $s(L)\sim0.01$ so that the signal field does not experience this low saturation regime. For the latter case  (see Fig. \ref{Switch}b), we expect to have better storage efficiencies when the storage step begins sharply so that atomic quantities --in particular the first-order population difference-- cannot follow the leaking signal while being absorbed. Moreover, we expect the sharpness of the retrieval step to only weakly affect the storage efficiency because we actually want the retrieval to leak.

